\documentstyle[seceq,epsfig]{ptptex}

\title{Two-Dimensional Tensor Product Variational Formulation}

\author
{Tomotoshi {\sc Nishino},\footnote{E-mail address:
nishino@phys.sci.kobe-u.ac.jp} Yasuhiro {\sc Hieida},
Kouichi {\sc Okunishi}$^{1,2}_{~}$,
Nobuya {\sc Maeshima}$^{3}_{~}$,
Yasuhiro {\sc Akutsu}$^{3}_{~}$,
and  Andrej {\sc Gendiar}$^{4}_{~}$ }

\inst
{Department of Physics, Graduate School of Science,
Kobe University, 657-8501\\
$^{1}_{~}$ Department of Applied Physics, Graduate School of Engineering,
Osaka University, 565-0781\\
$^{2}_{~}$ Department of Physics, Faculty of Science,
Niigata University, 950-2181\\
$^{3}_{~}$ Department of Physics, Graduate School of Science,
Osaka University, 560-0043\\
$^{4}_{~}$ Institute of Physics, Slovak Academy of Sciences,
D\'ubravsk\'a cesta 9, SK-842 28 Bratislava, SLOVAKIA}

\recdate{}

\abst{
We propose a numerical self-consistent method for 3D classical
lattice models, which optimizes the variational state written as
two-dimensional product of tensors. The variational partition
function is calculated by the corner transfer matrix renormalization
group (CTMRG), which is a variant of the density matrix renormalization
group (DMRG). Numerical efficiency of the method is observed via its 
application to the 3D Ising model.
}

\begin{document}

\maketitle

\section{Introduction}

Variational states represented as  products of local tensols have been
used widely for calculating the ground-state energy in  one-dimensional 
(1D) quantum systems and the free energy in two-dimensional (2D) 
classical statistical systems. In 1941 Kramers and Wannier calculated 
a lower bound of the partition function of the 2D Ising model using the 
matrix product state,~\cite{Krm} which is a prototype of the tensor 
product state (TPS). A remarkable point in their variational approach 
is that the calculated lower bound is a good approximation for the 
exact partition function~\cite{Onsager} especially in low and high 
temperature regions. About 30 years later Baxter formulated a 
variational method, which is a natural extension of the Kramers-Wannier 
approximation, by expressing the eigenstate of the transfer matrix in 
the form of TPS.~\cite{Bx1,Bx2,Bx3}He has also shown that the 
integrable model can be solved using the corner transfer matrix 
generated from a local tensor with infinite degree of freedom.
Quite recently the density matrix renormalization group 
(DMRG)~\cite{Wh1,Wh2,rev} has provided a very practical numerical 
procedure to perform the variational computation for the 
position-dependent TPS.~\cite{Fannes1,Fannes2,Ostlund}

A current interest about TPS is its application to higher dimensional 
systems.~\cite{rev,Sierra1,CTTRG} For example, Niggemann {\it et al} 
has shown that the ground state of the 2D valence-bond-solid (VBS) type 
quantum spin model is exactly expressed as the TPS, where the 
calculation of the ground-state expectation value is reduced to that of 
the partition function for the corresponding 2D vertex 
model.~\cite{Zitt,Zitt2} Hieida {\it et al} evaluated the correlation 
function of the 2D deformed VBS model by combining DMRG with 
the {\it exact} variational formulation of the system.~\cite{Hieida}
Also for the 3D classical models, the authors have been  generalizing 
the 2D TPS, in the context of the higher-dimensional 
DMRG.~\cite{CTTRG,kW3D,TPVA}

In constructing  the TPS, the most important step is how to represent 
the local tensors,  since it often reflects on the efficiency of the 
variational calculation. In our previous works, we calculated the variational 
free energy of the 3D Ising model, employing the interaction-round-a-face 
(IRF) representation of the local tensor with 16 variational 
parameters.~\cite{TPVA} Such a variational state is often called as 
the IRF-type TPS.~\cite{Sierra1,TPVA} The resulting efficiency is quite 
good in the off-critical region, though the calculated transition 
temperature is about $1.5\%$ higher than that obtained by the Monte 
Carlo (MC) simulations.~\cite{MC1,MC2} However, we have only discussed 
the efficiency of the IRF-type TPS and thus it is necessary to challenge 
another type of TPS.

In this paper, we construct the vertex-type TPS in 2 dimension, by 
introducing four auxiliary variables of $m$-states into a local tensor.
The new variational state has $2m^4_{~}$ parameters, which is twice 
as many as that of the IRF-type TPS when $m = 2$. Numerical efficiency 
of this TPS is examined through its application to a 3D vertex model 
whose thermodynamic property is equivalent to the 3D Ising 
model.~\cite{CTMRG2} In the next section, we show the construction 
of the 2D TPS with auxiliary variables. Numerical procedures to obtain 
the maximum of the variational partition function is shown in \S 3, 
and the calculated result is explained in \S 4.
Conclusions are summarized in \S 5.

\section{Variational Formulation}

As an example of the 3D classical model, we consider the 3D vertex 
model that has one-to-one correspondence with the simple cubic lattice 
Ising model.~\cite{CTMRG2} Let us start from a brief description of 
this vertex model. Consider a simple cubic lattice of the size 
$N \times N \times \infty$ in the XYZ-directions, where a 2-state 
spin variable that takes either up ($+$) or down ($-$) are 
sitting on each link between the nearest lattice points. 
Thus one lattice point is surrounded by 6 spins as shown in Fig.~1.
This is the unit of the 3D vertex model, which is called a `vertex'.
We label the spins on the vertical links of the vertex as $s$ (bottom)
and $\bar s$ (top), and those on the horizontal links as $\sigma$,
$\sigma'$, $\sigma''$, and $\sigma'''$.

Statistical property of the vertex model is specified by the Boltzmann
weight $w_{{\bar s} s}^{~}( \sigma \, \sigma' \sigma'' \sigma''' )$
assigned to the vertex. In the following, we set the weight as
\begin{equation}
w_{{\bar s} s}^{~}( \sigma \, \sigma' \sigma'' \sigma''' )
= \sum_{x = \pm 1}^{~}
U_{\bar s}^{x} \, U_{s}^{x} \,
U_{\sigma}^{x} \, U_{\sigma'}^{x} \,
U_{\sigma''}^{x} \, U_{\sigma'''}^{x} \, ,
\end{equation}
where $U_{y}^{x}$ is unity when $x = y$ and is
$e^{\beta J}_{~} + \sqrt{e^{2{\beta J}}_{~}-1}$ otherwise;
the 3D vertex model has the same free energy as the 3D Ising model 
with the Hamiltonian ${\cal H}=\sum J x x'$, where  
$x$ and $x'$ denote the neighboring Ising spin variables.~\cite{CTMRG2}

\begin{figure}
\epsfxsize=100mm
\centerline{\epsffile{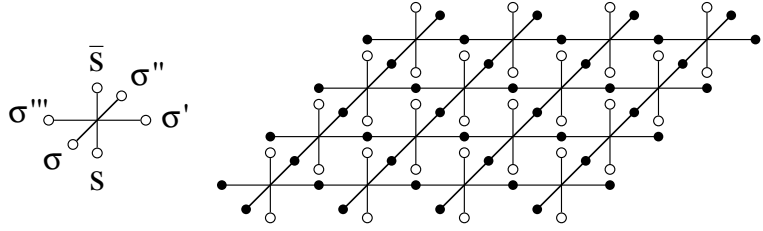}}
\caption{ Spin variables around a vertex, and the
transfer matrix of the 3D vertex model. }
\label{fig1}
\end{figure}

Let us introduce two notations in order to simplify the following
expressions. The first one is $\{ \sigma \}$, which represents the 
group of horizontal spins $\sigma, \sigma', \sigma''$, and 
$\sigma'''$ around a vertex. For example, the vertex
weight $w_{{\bar s} s}^{~}( \sigma \, \sigma' \sigma'' \sigma''' )$
is simply written as $w_{{\bar s} s}^{~}\{ \sigma \}$. The second
one is the matrix representation of the Boltzmann weight
\begin{equation}
W\{ \sigma \} \equiv
\left( \begin{array}{cc}
w_{++}^{~}\{ \sigma \} &  w_{+-}^{~}\{ \sigma \} \\
w_{-+}^{~}\{ \sigma \} &  w_{--}^{~}\{ \sigma \}
\end{array} \right) \, ,
\end{equation}
where we have regarded $s$ and $\bar s$ of
$w_{{\bar s} s}^{~}\{ \sigma \}$, respectively, as column and row
indices of the $2 \times 2$ matrix.

The system explained above can be interpreted as the infinite 
stack of $N \times N$ layers, where each layer plays the role of 
the layer-to-layer transfer matrix $T$. Using the matrix expression 
in Eq.~(2.2) and writing the vertex weight at the position $( i, j )$ in 
the layer as $W\{ \sigma_{ij}^{~} \}$, we can express the 
transfer matrix as a two-dimensionally connected vertices:
\begin{equation}
T \equiv \sum_{[ \sigma ]}^{~} \prod_{1 \le ij \le N}^{~}
W\{ \sigma_{ij}^{~} \} \, ,
\end{equation}
where $\sum_{[ \sigma ]}^{~}$ denotes the configuration sum for all 
the spins on the horizontal links, that are shown by black circles in Fig.~1.
(We use black marks for the spin variables whose configuration sums are taken.)
Our interest is to calculate the variational partition function per layer
\begin{equation}
\lambda\mbox{\boldmath $[$} \Psi \mbox{\boldmath $]$} =
\frac{\langle \Psi | T | \Psi \rangle}{\langle \Psi | \Psi \rangle}
\label{rratio}
\end{equation}
for a given TPS $| \Psi \rangle$, and further to find out the best 
TPS that minimizes 
$\lambda\mbox{\boldmath $[$} \Psi \mbox{\boldmath $]$}$.

\begin{figure}
\epsfxsize=100mm
\centerline{\epsffile{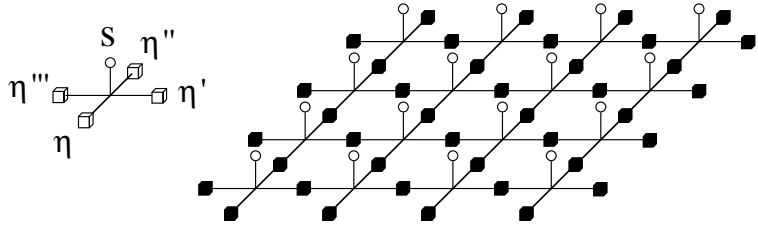}}
\caption{ Spin variables around a variational tensor, and the
variational state constructed as 2D product of the tensors. }
\label{fig2}
\end{figure}

Now we explicitly construct the trial eigenvector $| \Psi \rangle$ as
the 2D tensor product, as is shown in Figure 2.~\cite{Sierra1,TPVA}
Each tensor has an apex spin $s = \pm 1$ and four $m$-state auxiliary
variables $\eta$, $\eta'$, $\eta''$, and $\eta'''$ shown by cubes.
Let us express the elements of the tensor
as $v_{~}^{s}\{ \eta \}$, where $\{ \eta \}$ denotes the set of 
auxiliary variables $\eta, \eta', \eta''$, and $\eta'''$.
As we have expressed the vertex weight as a matrix in Eq.~(2.2), 
it is useful to interpret $v_{~}^{+}\{ \eta \}$ and $v_{~}^{-}\{ \eta \}$ 
as components of the column vector
\begin{equation}
{\bf V}\{ \eta \} \equiv
\left( \begin{array}{c}
v_{~}^{+}\{ \eta \} \\
v_{~}^{-}\{ \eta \}
\end{array} \right) \, .
\end{equation}
Using this notation, we construct the 2D TPS as
\begin{equation}
| \Psi \rangle \equiv \sum_{[ \eta ]}^{~} \prod_{1 \le ij \le N}^{~}
{\bf V}\{ \eta_{ij}^{~} \} \, , \label{vtps}
\end{equation}
where $\{ \eta_{ij}^{~} \}$ denotes the set of auxiliary variables 
around the lattice point $( i, j )$, and the configuration sum 
$\sum_{[ \eta ]}^{~}$ is taken over for all the auxiliary 
variables shown by black cubes in Fig.~2. Since the above construction 
of the TPS is similar to that of the transfer matrix $T$ of the 3D vertex 
model, we call the TPS (\ref{vtps}) as the {\it vertex-type} 
TPS.~\cite{Sierra1,TPVA}

\begin{figure}
\epsfxsize=62mm
\centerline{\epsffile{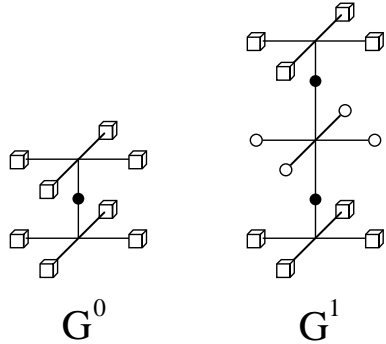}}
\caption{ The vertex weight $G^0_{~}$ of the 2-layer classical
system (Eq.~(2.9)), and $G^1_{~}$ of the 3-layer one (Eq.~(2.10)).}
\label{fig3}
\end{figure}

The main profit of expressing the variational state in the form of 
the TPS is that both $\langle \Psi | \Psi \rangle$ and 
$\langle \Psi | T | \Psi \rangle$ can be expressed as the partition 
functions of 2- and 3-layer 2D
classical lattice models, respectively.~\cite{Zitt,Hieida,kW3D,TPVA}
Let us consider $\langle \Psi | \Psi \rangle$ first.
 From the definition of $| \Psi \rangle$ in Eq.~(2.6) we obtain
\begin{equation}
\langle \Psi | \Psi \rangle = \sum_{[ \eta ]\,[ \bar \eta ]}^{~}
\prod_{1 \le ij \le N}^{~}
\,\, {\bf V}\{ {\bar \eta}_{ij}^{~} \} \cdot {\bf V}\{ \eta_{ij}^{~} \}
= \sum_{[ \eta ]\,[ \bar \eta ]}^{~} \prod_{1 \le ij \le N}^{~}
\left( \sum_{s=\pm 1}^{~}
v_{~}^s\{ {\bar \eta}_{ij}^{~} \} v_{~}^s\{ \eta_{ij}^{~} \} \right) \, ,
\end{equation}
where the configuration sum is taken over for all the spins and the
auxiliary variables. Introducing the new notation
\begin{equation}
G^0_{~}\{ {\bar \eta}_{ij}^{~} | \eta_{ij}^{~} \} \equiv
{\bf V}\{ {\bar \eta}_{ij}^{~} \} \cdot {\bf V}\{ \eta_{ij}^{~} \}
= \sum_{s=\pm 1}^{~}
v_{~}^s\{ {\bar \eta}_{ij}^{~} \} v_{~}^s\{ \eta_{ij}^{~} \}
\end{equation}
and regarding it as a local Boltzmann weight,
$\langle \Psi | \Psi \rangle$ can be interpreted as the
partition function of a 2-layer 2D vertex model
\begin{equation}
Z^0_{~} \equiv \sum_{[ \eta ] \, [ \bar \eta ]}^{~} \prod_{1 \le ij \le N}^{~}
\, G^0_{~}\{ {\bar \eta}_{ij}^{~} | \eta_{ij}^{~} \} \, .
\end{equation}
(See Fig.~3.)
In the same manner, $\langle \Psi | T | \Psi \rangle$ can be expressed as the
partition function of a 3-layer 2D vertex model:
\begin{equation}
Z^1_{~}
\equiv \sum_{[ \eta ] \, [ \sigma ] \, [ \bar \eta ]}^{~} \prod_{1 \le ij 
\le N}^{~}
\, G^1_{~}\{ {\bar \eta}_{ij}^{~} | \sigma_{ij}^{~} | \eta_{ij}^{~} \} ,
\end{equation}
which is characterized by the local Boltzmann weight:
\begin{equation}
G^1_{~}\{ {\bar \eta}_{ij}^{~} | \sigma_{ij}^{~} | \eta_{ij}^{~} \} \equiv
{\bf V}\{ {\bar \eta}_{ij}^{~} \} \cdot
\left(W\{ \sigma_{ij}^{~} \}  {\bf V}\{ \eta_{ij}^{~} \}\right )
= \sum_{{\bar s} s}^{~} v_{~}^{\bar s}\{ {\bar \eta}_{ij}^{~} \}
w_{{\bar s} s}^{~}\{ \sigma_{ij}^{~} \}
v_{~}^s\{ \eta_{ij}^{~} \} \, .
\end{equation}

Thus once the $2m^4_{~}$ numbers of the elements (of $v_{~}^s\{ \eta \}$) 
are given, the variational ratio 
$\lambda\mbox{\boldmath $[$} \Psi \mbox{\boldmath $]$} = Z^1_{~} / Z^0_{~}$ 
is easily obtained via standard numerical methods for the 2D classical 
lattice models. What we have to consider next is the way to find out 
the best tensor $v_{~}^s\{ \eta \}$ which 
maximizes $\lambda\mbox{\boldmath $[$} \Psi \mbox{\boldmath $]$}$.

\section{Maximization of the Variational Partition Function}

In order to maximize the variational partition function
$\lambda\mbox{\boldmath $[$} \Psi \mbox{\boldmath $]$}$,
we consider its variation:
\begin{equation}
\delta \lambda\mbox{\boldmath $[$} \Psi \mbox{\boldmath $]$}
\equiv \sum_{ij}^{~} \,
  \frac{\partial \lambda\mbox{\boldmath $[$} \Psi \mbox{\boldmath $]$}}
  {\partial v_{~}^{\bar s}\{ {\bar \eta}_{ij}^{~} \}}
  \delta v_{~}^{\bar s}\{ {\bar \eta}_{ij}^{~} \}
+ \sum_{ij}^{~} \,
\frac{\partial \lambda\mbox{\boldmath $[$} \Psi \mbox{\boldmath $]$}}
{\partial v_{~}^{s}\{ \eta_{ij}^{~} \}}
\delta v_{~}^{s}\{ \eta_{ij}^{~} \}
\end{equation}
with respect to the infinitesimal change of the local tensors
\begin{eqnarray}
v_{~}^{\bar s}\{ {\bar \eta}_{ij}^{~} \} &\rightarrow&
v_{~}^{\bar s}\{ {\bar \eta}_{ij}^{~} \} +
\delta v_{~}^{\bar s}\{ {\bar \eta}_{ij}^{~} \} \nonumber\\
v_{~}^{s}\{ \eta_{ij}^{~} \} &\rightarrow&
v_{~}^{s}\{ \eta_{ij}^{~} \} +
\delta v_{~}^{s}\{ \eta_{ij}^{~} \} \, .
\end{eqnarray}
When the linear dimension of the system ($= N$) is sufficiently 
large, most of the terms in the r.h.s. of Eq.~(3.1) are very close 
with each other. This is  because the boundary is far away from 
most of the sites and hence the system can be regarded to be 
uniform in the thermodynamic limit. Thus it is sufficient to treat 
the variation of $\lambda\mbox{\boldmath $[$} \Psi \mbox{\boldmath 
$]$}$ at the center of the system:
\begin{equation}
\delta_{\rm c}^{~} \lambda\mbox{\boldmath $[$} \Psi \mbox{\boldmath $]$}
\equiv \frac{\partial
\lambda\mbox{\boldmath $[$} \Psi \mbox{\boldmath $]$}}
{\partial v_{~}^{\bar s}\{{\bar \eta}_{\rm c}^{~} \}}
\delta v_{~}^{\bar s}\{{\bar \eta}_{\rm c}^{~} \}
+ \frac{\partial \lambda\mbox{\boldmath $[$} \Psi \mbox{\boldmath $]$}}
{\partial v_{~}^{s}\{ \eta_{\rm c}^{~} \}}
\delta v_{~}^{s}\{ \eta_{\rm c}^{~} \} \, ,
\end{equation}
where $v_{~}^{\bar s}\{ {\bar \eta}_{\rm c}^{~} \}$ and
$v_{~}^{s}\{ \eta_{\rm c}^{~} \}$ are tensors at the center 
${\rm c} = ( N/2, N/2 )$.

For the convenience of writing down $\delta_{\rm c}^{~}
\lambda\mbox{\boldmath $[$} \Psi \mbox{\boldmath $]$}$ more explicitly,
we divide $Z^0_{~} = \langle \Psi | \Psi \rangle$ and 
$Z^1_{~} = \langle \Psi | T | \Psi \rangle$  into two parts as follows:
\begin{eqnarray}
Z^0_{~} &=& \sum_{\{ {\bar \eta}_{\rm c}^{~} \} \,
\{ \eta_{\rm c}^{~} \} }^{~}
G^0_{~}\{ {\bar \eta}_{\rm c}^{~} | \eta_{\rm c}^{~} \} \,
X^0_{~}\{ {\bar \eta}_{\rm c}^{~} | \eta_{\rm c}^{~} \} \,
\nonumber\\
Z^1_{~} &=& \sum_{\{ {\bar \eta}_{\rm c}^{~} \} \,
\{ \sigma_{\rm c}^{~} \} \, \{ \eta_{\rm c}^{~} \} }^{~}
G^1_{~}\{ {\bar \eta}_{\rm c}^{~} | \sigma_{\rm c}^{~} |
\eta_{\rm c}^{~} \} \,
X^1_{~}\{ {\bar \eta}_{\rm c}^{~} | \sigma_{\rm c}^{~} |
\eta_{\rm c}^{~} \} \, .
\end{eqnarray}
The factors $G^{0}_{~}$ and $G^{1}_{~}$ are the ``local Boltzmann weights'' 
at the center of the system,  and  
$X^0_{~}\{ {\bar \eta}_{\rm c}^{~} | \eta_{\rm c}^{~} \}$ and 
$X^1_{~}\{ {\bar \eta}_{\rm c}^{~} | \sigma_{\rm c}^{~} | \eta_{\rm c}^{~} \}$ 
are the rest of the system. The new tensors  
$X^0_{~}\{ {\bar \eta}_{\rm c}^{~} | \eta_{\rm c}^{~} \}$ and 
$X^1_{~}\{ {\bar \eta}_{\rm c}^{~} | \sigma_{\rm c}^{~} | \eta_{\rm c}^{~} \}$  
play a role of ``reservoirs'' in 
the terminology of the DMRG, which are defined as
\begin{eqnarray}
X^0_{~}\{ {\bar \eta}_{\rm c}^{~} | \eta_{\rm c}^{~} \}
&\equiv&
\sum_{[ \bar \eta ]' \, [ \eta ]'}^{~}
\prod_{(ij) \neq {\rm c}}^{~}
G^0_{~}\{ \bar \eta_{ij}^{~} | \eta_{ij}^{~} \} \nonumber\\
X^1_{~}\{ {\bar \eta}_{\rm c}^{~} | \sigma_{\rm c}^{~} |
\eta_{\rm c}^{~} \}
&\equiv&
\sum_{[ \bar \eta ]' \, [ \sigma ]' \, [ \eta ]'}^{~}
\prod_{(ij) \neq {\rm c}}^{~}
G^1_{~}\{ \bar \eta_{ij}^{~} | \sigma_{ij}^{~} | \eta_{ij}^{~} \} \, ,
\end{eqnarray}
where the restricted product $\prod_{(ij) \neq {\rm c}}^{~}$ denotes that
$G^0_{~}\{ {\bar \eta}_{\rm c}^{~} | \eta_{\rm c}^{~} \}$ and
$G^1_{~}\{ \bar \eta_{\rm c}^{~} | \sigma_{\rm c}^{~} | \eta_{\rm c}^{~} \}$
are not included in the right hand sides, and the restricted sum 
$\sum_{[ \bar \eta ]' \, [ \eta ]'}^{~}$ and
$\sum_{[ \bar \eta ]' \, [ \sigma ]' \, [ \eta ]'}^{~}$ denote spin 
configuration sum for all the spins except for those at the center, 
${\bar \eta_{\rm c}^{~}}$, $\sigma_{\rm c}^{~}$, and 
$\eta_{\rm c}^{~}$. Thus the division by Eq.~(3.4) is equivalent to 
punch out the system a the center; this operation is similar to puncture 
the system in the `puncture renormalization group' by Mart\'{\i}n-Delgado
{\it et al}.~\cite{PunctureRG}

Substituting the definitions of $G^0$ and $G^1$ into Eq.~(3.4) and  
introducing two matrices
\begin{eqnarray}
A_{{\bar s} s}^{~}\{ {\bar \eta}_{\rm c}^{~} | \eta_{\rm c}^{~} \}
&\equiv& ~~~~~\,
X^0_{~}\{ {\bar \eta}_{\rm c}^{~} | \eta_{\rm c}^{~} \} \,
\delta_{{\bar s} s}^{~} \nonumber\\
B_{{\bar s} s}^{~}\{ {\bar \eta}_{\rm c}^{~} | \eta_{\rm c}^{~} \}
&\equiv& \sum_{\{ \sigma_{\rm c}^{~} \} }^{~}
X^1_{~}\{ {\bar \eta}_{\rm c}^{~} | \sigma_{\rm c}^{~} |
\eta_{\rm c}^{~} \} \,
w_{{\bar s} s}^{~}\{ \sigma_{\rm c}^{~} \}
\end{eqnarray}
we can express $Z^0_{~}$ and $Z^1_{~}$ in the binary form
\begin{eqnarray}
Z^0_{~} =
{\bf V}_{\rm c}^{\rm T} A \, {\bf V}_{\rm c}^{~} &\equiv&
\sum_{{\bar s} \, \{ {\bar \eta}_{\rm c}^{~} \}}^{~}
\sum_{s \, \{ \eta_{\rm c}^{~} \}}^{~}
v^{\bar s}_{~}\{ \eta_{\rm c}^{~} \}
A_{{\bar s} s}^{~}\{ {\bar \eta}_{\rm c}^{~} | \eta_{\rm c}^{~} \}
v^s_{~}\{ \eta_{\rm c}^{~} \} \nonumber\\
Z^1_{~} =
{\bf V}_{\rm c}^{\rm T} B \, {\bf V}_{\rm c}^{~} &\equiv&
\sum_{{\bar s} \, \{ {\bar \eta}_{\rm c}^{~} \}}^{~}
\sum_{s \, \{ \eta_{\rm c}^{~} \}}^{~}
v^{\bar s}_{~}\{ \eta_{\rm c}^{~} \}
B_{{\bar s} s}^{~}\{ {\bar \eta}_{\rm c}^{~} | \eta_{\rm c}^{~} \}
v^s_{~}\{ \eta_{\rm c}^{~} \}
\end{eqnarray}
of the $2m^4_{~}$-dimensional vector ${\bf V}_{\rm c}^{~}$.
Substituting the above expressions into Eq.~(3.3), we finally obtain the
stationary condition $\delta_{\rm c}^{~}
\lambda\mbox{\boldmath $[$} \Psi \mbox{\boldmath $]$} = 0$
expressed as a $2m^4_{~}$-dimensional generalized eigenvalue problem
\begin{equation}
\sum_{s \, \{ \eta_{\rm c}^{~} \}}^{~}
B_{{\bar s} s}^{~}\{ {\bar \eta}_{\rm c}^{~} | \eta_{\rm c}^{~} \}
v^s_{~}\{ \eta_{\rm c}^{~} \}
= \lambda\mbox{\boldmath $[$} \Psi \mbox{\boldmath $]$}
\, \sum_{s \, \{ \eta_{\rm c}^{~} \}}^{~}
A_{{\bar s} s}^{~}\{ {\bar \eta}_{\rm c}^{~} | \eta_{\rm c}^{~} \}
v^s_{~}\{ \eta_{\rm c}^{~} \} \, ,
\end{equation}
which can be abbreviated as
$B \, {\bf V}_{\rm c}^{~} =
\lambda\mbox{\boldmath $[$} \Psi \mbox{\boldmath $]$}
\, A \, {\bf V}_{\rm c}^{~}$.

This is a non-linear equation for the tensors $v^{\bar s}_{~}\{\bar \eta \}$
and $v^s_{~}\{ \eta \}$, because the `matrices' $A$ and $B$ themselves
are constructed from the tensors.
Therefore, equation (3.8) should be interpreted as a self-consistent
relation for the local tensors. A way to find out the solution of Eq.~(3.8)
is to repeat numerical substitutions as follows:
\vskip 3mm
\begin{itemize}
\item[(a)] Set a certain initial value to the tensor ${\bf V}$, which has
$2m^4_{~}$ elements.
\item[(b)] Numerically calculate  $X^0_{~}$ and $X^1_{~}$ by Eqs.~(3.5).
This calculation can be easily done with the help of the corner transfer
matrix renormalization group (CTMRG).~\cite{TPVA,CTMRG2,CTMRG1}
\item[(c)] Create $A$ and $B$ by Eqs.~(3.6), and apply $A^{-1}_{~}B$
to the tensor to obtain ${\bf V}' = A^{-1}_{~}B \, {\bf V}$.
\item[(d)] Create a new variational tensor ${\bf V} + \epsilon {\bf V}'$ where
$\epsilon$ is a small parameter. (We set $\epsilon = 0.1$.) Use
${\bf V} + \epsilon {\bf V}'$ as the new variational tensor and return to (b).
\item[(e)] Stop the calculation when
$\lambda\mbox{\boldmath $[$} \Psi \mbox{\boldmath $]$}$
does not increase any more.
\end{itemize}
\vskip 3mm
This numerical iteration works when the matrix $A$ is
regular and positive definite.
This condition is at least satisfied in the neighborhood of the stationary
point where $\lambda\mbox{\boldmath $[$} \Psi \mbox{\boldmath $]$}$
takes its maximum, but is not in general for arbitrary ${\bf V}$.
Thus, the initial choice of the variational tensor ${\bf V}$ is relevant to the
stability of the numerical calculation.

\section{Numerical Result}

We perform a calculation for the simplest vertex-type TPS with $m = 2$.
Note that each tensor has $2m^4_{~}=32$ elements, which is twice as 
many as that of the IRF-type TPS.~\cite{TPVA}
To start the numerical calculation, we set the initial tensor
$v^s_{~}\{ \eta \}$ from the vertex weight
\begin{equation}
v^s_{~}( \eta \, \eta' \eta'' \eta''' )  =
w_{+ s}^{~}( \sigma \, \sigma' \sigma'' \sigma''' ) + a \,
w_{- s}^{~}( \sigma \, \sigma' \sigma'' \sigma''' )
\end{equation}
where $a \sim 1$ is a constant that weakly breaks the spin
inversion symmetry $s \rightarrow -s$ of the initial TPS;
typically we set $a = 1.01$.
The construction of the  initial tensor explicitly uses the fact that $m = 2$.
We then performed the numerical self-consistent improvement for
$v^s_{~}\{ \eta \}$, and have succeeded to reach the fixed
point that satisfies Eq.~(3.8) within 1000 iterations in the
whole temperature regions.

In order to check the quality of the obtained TPS, we observe
the spontaneous magnetization of the 3D Ising model. In the
vertex representation of the 3D Ising model (Eq.~(2.1)), the
magnetization with respect to the optimized TPS is expressed as
\begin{equation}
M = \,
\frac{{\bf V}_{\rm c}^{\rm T} O \, {\bf V}_{\rm c}^{~}}
{ {\bf V}_{\rm c}^{\rm T} A \, {\bf V}_{\rm c}^{~}}
\left( \frac{{\bf V}_{\rm c}^{\rm T} B \, {\bf V}_{\rm c}^{~}}
{ {\bf V}_{\rm c}^{\rm T} A \, {\bf V}_{\rm c}^{~}} \right)^{-1} \!\!\!
= \,
\frac{{\bf V}_{\rm c}^{\rm T} O \, {\bf V}_{\rm c}^{~}}
{ {\bf V}_{\rm c}^{\rm T} B \, {\bf V}_{\rm c}^{~}} \, ,
\end{equation}
where the new matrix $O$ is --- similar to the matrix $B$ --- defined as
\begin{equation}
O_{{\bar s} s}^{~}\{ {\bar \eta}_{\rm c}^{~} | \eta_{\rm c}^{~} \}
\equiv \sum_{\{ \sigma_{\rm c}^{~} \} }^{~}
X^1_{~}\{ {\bar \eta}_{\rm c}^{~} | \sigma_{\rm c}^{~} | \eta_{\rm c}^{~} \} \,
o_{{\bar s} s}^{~}\{ \sigma_{\rm c}^{~} \} \,
\end{equation}
with the modified vertex weight
\begin{equation}
o_{{\bar s} s}^{~}\{ \sigma \} = \sum_{x = \pm 1}^{~} \, x \,
U_{\bar s}^{x} \, U_{s}^{x} \,
U_{\sigma}^{x} \, U_{\sigma'}^{x} \,
U_{\sigma''}^{x} \, U_{\sigma'''}^{x} ,
\end{equation}
which represents polarization of the Ising spin at the center of the system.

\begin{figure}
\epsfxsize=92mm
\centerline{\epsffile{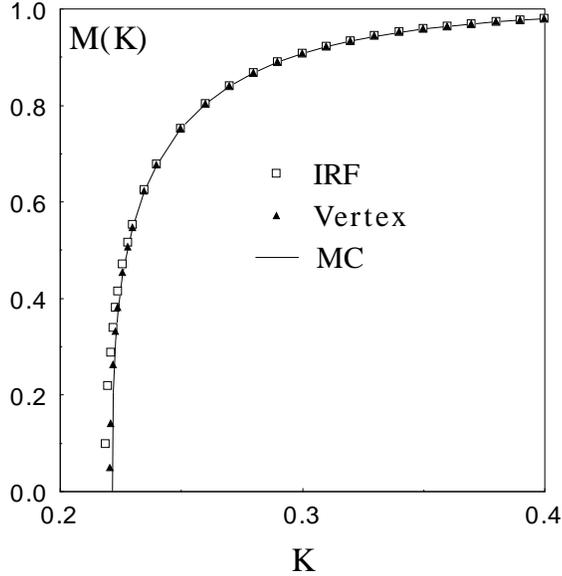}}
\caption{ Spontaneous magnetization $M^{\rm V}_{~}(K)$ of the 3D Ising
model calculated via TPS when $m = 2$. (See black triangles.)
For comparison, magnetization obtained by the IRF-type variational
calculation~\cite{TPVA} $M^{\rm IRF}_{~}(K)$ (white squares)
and that by a Monte Carlo simulation~\cite{MC1}
$M^{\rm MC}_{~}(K)$ (curve) are shown. }
\label{figX}
\end{figure}

We show the calculated result in figure 4.
The black triangles  indicates the calculated spontaneous magnetization 
$M^{\rm V}_{~}(K \equiv \beta J)$ by the vertex-type TPS.
For comparison, we also show the magnetization $M^{\rm IRF}_{~}(K)$ 
calculated by the IRF-type variational formulation~\cite{TPVA} 
(white squares) and the Monte Carlo result $M^{\rm MC}_{~}(K)$ 
(line) by Talapov and Bl\"ote.~\cite{MC1} Away from the critical 
point,  $M^{\rm V}_{~}(K)$ shows good agreement with $M^{\rm 
MC}_{~}(K)$, but as $K$ approaches to the critical point $M^{\rm V}_{~}(K)$ 
deviates from $M^{\rm MC}_{~}(K)$. The estimated critical point by 
the vertex-type TPS is $K_{\rm c}^{\rm V} = 0.2203$, which is 
$0.6\%$ smaller than $K_{\rm c}^{\rm MC} = 0.2216544$~\cite{MC1,MC2}.
We can see that the result for the vertex-type TPS is better than that 
for the IRF-type variational formulation $K_{\rm c}^{\rm IRF} = 0.2188$, 
which is $1.5\%$ smaller than $K_{\rm c}^{\rm MC}$. Thus we can 
conclude that vertex-type TPS is more efficient than the IRF-type 
TPS.~\cite{TPVA} We note that the observed critical behavior of 
$M^{\rm V}_{~}(K)$ in the vicinity of the critical point is  the mean-field 
one $M^{\rm V}_{~}(K) \propto \sqrt{K - K_{\rm c}^{V}}$.
It should be remarked that the numerical computation time  with the 
vertex formulation is the same order as that of the IRF one.

\section{Conclusions and Discussions}

We have proposed a numerical self-consistent method for 3D
classical systems, which optimizes the 2D vertex-type TPS
with $2m^4_{~}$ variational parameters. The method is applied 
to the 3D vertex model, which has the same thermodynamic 
property with the 3D Ising model, and we have confirmed that 
the vertex-type TPS gives better transition temperature than 
the IRF-type TPS.

If we increase the number of states $m$ of the auxiliary variables 
of the vertex-type TPS we will be able to further tune the vertex-type TPS.
It is possible to make a calculations of  $A$ and $B$ in Eq.~(3.6) 
with a short computation of a few  seconds up to $m \sim 4$.
We have, however, not succeeded in finding out the solution 
of the self-consistent equation (Eq.~(3.8)) stably for the cases $m > 2$. 
This is because we encounter a numerical problem that the matrix $A$ 
becomes singular during the self-consistent calculation. It is our future 
problem to find out a systematic way to set up the initial tensor so 
that $A$ is always regular and positive definite.

\section*{Acknowledgments}

T.~N. thank to G.~Sierra and M.A.~Mart\'{\i}n-Delgado
for valuable discussions about tensor product formulations and the puncture 
renormalization group.
This work was partially supported by the ``Research
for the Future'' Program from The  Japan Society for the Promotion of Science
(JSPS-RFTF97P00201) and by the Grant-in-Aid for  Scientific Research from
Ministry of Education, Science, Sports and Culture (No.~09640462 and
No.~11640376). Most of the numerical calculations were
performed by Compaq Fortran
on the HPC-Alpha UP21264 Linux workstation.

\end{document}